\documentclass[3p,times,twocolumn]{elsarticle}

\usepackage{ecrc}

\volume{00}

\firstpage{1}

\journalname{Nuclear and Particle Physics Proceedings}

\runauth{}

\jid{nppp}

\jnltitlelogo{Nuclear and Particle Physics Proceedings}

\usepackage{amssymb}
\usepackage{amsmath}

\usepackage[figuresright]{rotating}

\usepackage[bookmarks]{hyperref}

\newcommand{\Dzero} {\mbox{$\mathrm D^{0}$}}
\newcommand{\Dplus} {\mbox{$\mathrm D^{+}$}}
\newcommand{\Dstar} {\mbox{$\mathrm D^{*+}$}}
\newcommand{\Ds}    {\mbox{$\mathrm D^{+}_{s}$}}
\newcommand{\Jpsi}  {\mbox{$\mathrm{ J/\psi }$}}
\newcommand{\sqrtS} {\mbox{$\sqrt{s}$}}
\newcommand{\pt}    {\mbox{$p_{\text{\textsc{t}}}$}}
\newcommand{\gmom}  {\mbox{GeV/$c$}}

\newcommand{\DzeroDecayChan} {\mbox{$\mathrm{ D^{0} \rightarrow K^{-}\pi^{+}}$}}
\newcommand{\DplusDecayChan} {\mbox{$\mathrm{ D^{+} \rightarrow K^{-}\pi^{+}\pi^{+}}$}}
\newcommand{\DstarDecayChan} {\mbox{$\mathrm{ D^{*+}\rightarrow D^{0}\pi^{+}}$}}
\newcommand{\DsDecayChan}    {\mbox{$\mathrm{ D^{+}_{s} \rightarrow \phi\pi^{+} \rightarrow K^{+}K^{-}\pi^{+}}$}}

\begin{document}

\begin{frontmatter}



\dochead{}

\title{D-meson production in proton-proton collisions with ALICE at the LHC}


\author[label_univ]{Julien Hamon}
\author{for the ALICE Collaboration}
\address[label_univ]{Universit\'e de Strasbourg, CNRS-IN2P3, IPHC UMR 7178, F-67000 Strasbourg, France}

\begin{abstract}
    In this paper, an overview of recent $\Dzero$, $\Dplus$ and $\Dstar$ measurements, performed by ALICE
    in proton--proton collisions at $\sqrtS~=~2.76$, 5, 7, 8 and 13 TeV, is reported.
    The minimum-bias production cross sections, as well as their energy and species dependences,
    are compared to perturbative QCD calculations.
    The evolution of the D-meson yields with the event multiplicity
    is compared to phenomenological models.
\end{abstract}


\end{frontmatter}


    \section{Motivations}
    \label{sec:Motivations}

Heavy quarks (charm and beauty) are powerful probes to investigate the properties of the Quark-Gluon Plasma,
the hot and dense state of matter produced in ultra-relativistic heavy-ion collisions.
Because of their large masses, heavy quarks are produced in parton hard-scatterings,
occurring at the very beginning of hadronic collisions.
Thus, they participate in the whole evolution of the heavy-ion collision.

In addition to providing the essential baseline for nucleus-nucleus collision studies,
the study of production of heavy-flavour hadrons in proton-proton (pp) collisions allow
for studying also the perturbative sector of Quantum Chromo-Dynamics (pQCD).
Indeed, the large heavy quark masses (\mbox{$m_{c,b}\gg \Lambda_{\text{QCD}}$})
act as hard scales allowing the use of perturbation series in various kinematic regions.
In this context,
Fixed-Order-Next-to-Leading-Log (FONLL) \cite{fonll2012CharmAndBottomProductionAtLHC}
and General Mass Variable Flavour Number Scheme (GM-VFNS) \cite{gmvfns2012CharmProductionAtLHC}
represent the state-of-the-art of pQCD calculations of heavy-flavour hadron productions.
Both models implement, in two different ways, the standard collinear factorisation theorem at next-to-leading order (NLO).
Alternatively, calculations based on $k_{T}$-factorisation \cite{kTfactorisation2013CharmProductionAtLHC}
are performed at leading order in $\alpha_{s}$, including a part of real higher order corrections.

The measurements of heavy-flavour hadron productions at the Large Hadron Collider (LHC) energies
allow for testing those pQCD calculations down to zero transverse momentum ($\pt$).
Besides, differential measurements can help in studying more in detail
ingredients on which the pQCD calculations are based,
namely the parton distribution functions (PDF), the partonic hard scatterings and the fragmentation functions (FF).
As an example, the study of the species dependence of the production of heavy-flavour hadrons
allows the fragmentation functions to be investigated.
By comparing the heavy-flavour hadron productions in different rapidity regions and/or different energies,
pQCD models become sensitive to gluon PDF.
Partons with proton momentum fraction $x_B\sim10^{-5}$ might be accessed with LHC experiments
\cite{cacciari2015gluonPDFconstraintsFromRatioVsRapidityVsEnergy}.
Finally, several hard partonic interactions occurring in a single pp collision
are expected to play a role in the production of heavy quarks at LHC energies.
Such a phenomenon is usually referred as multi-parton interactions.
Multiplicity-dependent measurements of heavy-flavour hadron productions can shed light on
the relation between the production of charm hadrons and the event particle multiplicities.

    \section{Topological reconstruction of D mesons in ALICE}
    \label{sec:Reconstructions}

The ALICE experiment \cite{alice2008pstationExpmtInJINST,alice2014AlicePerformanceRunI}
has measured the production of charm and beauty mesons since the first data taking period of the LHC (Run~1).
Open-charm mesons $\Dzero$, $\Dplus$, $\Dstar$ and $\Ds$ have been studied
over a wide $\pt$ range,
typically from 1 to 24 $\gmom$, depending on the D-meson species and the colliding system.
The exclusive reconstruction of these four D mesons is performed at mid-rapidity ($|y|<0.5$)
via their hadronic decay channels:
$\DzeroDecayChan$ (with branching ratio, BR, of $3.93\pm0.04\%$),
$\DplusDecayChan$ (BR of $9.46\pm0.24\%$),
$\DstarDecayChan$ (BR of $67.7\pm0.5\%$) and
$\DsDecayChan$ (BR of $2.27\pm0.08\%$)
\cite{particleDataGroup2016}.
A detailed description of D-meson reconstruction in ALICE can be found in \cite{alice2012charmProductionInPP7TeV}.

The analysis of D mesons is based on the displacement of the reconstructed decay vertices with respect to the interaction vertex,
taking advantage of the D-meson decay length of few hundreds of $\mu$m.
Open-charm meson candidates are built up as pairs or triplets of tracks,
with the proper charge combination and taking into account particle identification
(from energy loss or time of flight information).
Kinematic and geometrical selections, such as the minimum distance between the decay and interaction vertices,
are then optimised to maximise the statistical significance of the signal per $\pt$ interval.
The raw yields of the reconstructed D mesons is
extracted with the help of a fit of invariant mass distributions.

The production cross section is evaluated from the raw yield
correcting for the detector acceptance (Acc) and efficiency ($\epsilon$),
by means of realistic Monte Carlo simulations,
which include the detector configuration and the LHC beam condition,
according to
\begin{equation}
    \left. \frac{\mathrm{d}^{2}\sigma^{\mathrm{D}}}{\mathrm{d}\pt\mathrm{d}y} \right|_{|y|<0.5}
    =
    \frac{\frac{1}{2} f_{\text{prompt}} \cdot \left. N^{\mathrm{D+\bar{D},raw}} \right|_{y<y_{\text{fid}}}}
    {(\text{Acc} \times \epsilon)_{\text{prompt}} \, \Delta\pt \, \Delta y}
    \,
    \frac{1}{L_{\text{int}}\, \text{BR}}
\end{equation}
where $\Delta y$ and $\Delta \pt$ are the rapidity and transverse momentum intervals,
$L_{\text{int}}$ is the integrated luminosity.
$f_{\text{prompt}}$ is the fraction of prompt D mesons present in the raw yield,
evaluated with theory-driven methods \cite{alice2017dmesonProductionInppAt7TeVpass4}.
The factor $\frac{1}{2}$ account for the fact that the raw yield contains particles and antiparticles.

    \section{Production cross section of D mesons in minimum-bias pp collisions}
    \label{sec:crossSectionPP13TeV}

The $\pt$-differential production cross sections of prompt $\Dzero$, $\Dplus$, $\Dstar$ and $\Ds$
have been measured at central rapidity by ALICE in minimum-bias pp collisions
collected during LHC Run~1, for various centre-of-mass energies:
$\sqrtS=2.76$ \cite{alice2012charmProdInPP2pt76TeV},
7 \cite{alice2012charmProductionInPP7TeV,alice2017dmesonProductionInppAt7TeVpass4,alice2016DmesonProductionInpPb5pt02TeVAndppAt7TeV}
and 8 TeV \cite{sharma2017pp8TeV}.

\begin{figure}[htb]
        \includegraphics[scale=0.38]{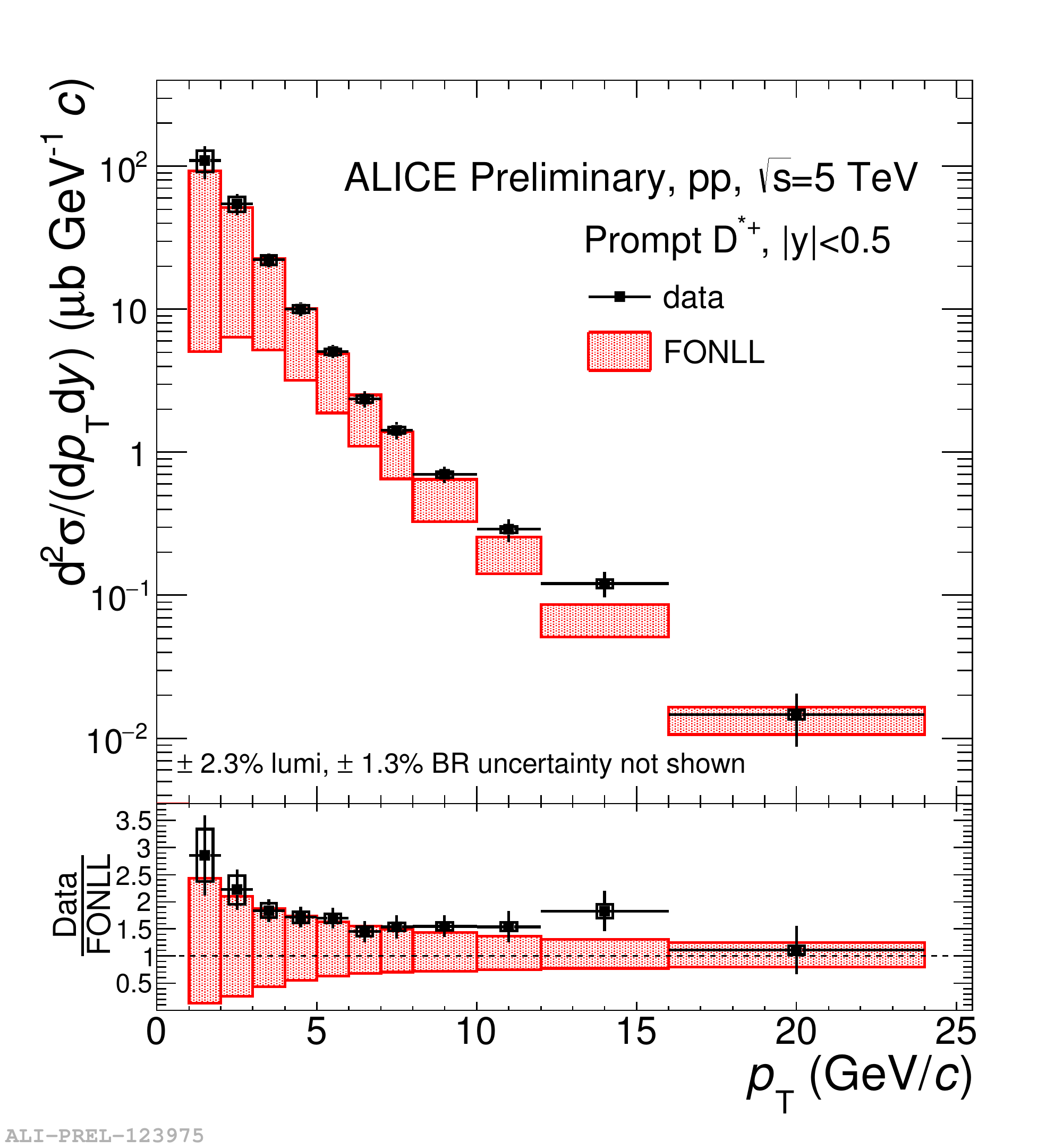}
        \caption{Preliminary $\pt$-differential production cross section of prompt $\Dstar$, measured by ALICE in pp collisions at $\sqrtS =$ 5 TeV,
        compared to FONLL prediction.}
        \label{fig:CrossSectionDstar}
    \end{figure}

\begin{figure}[htb]
        \includegraphics[scale=0.38]{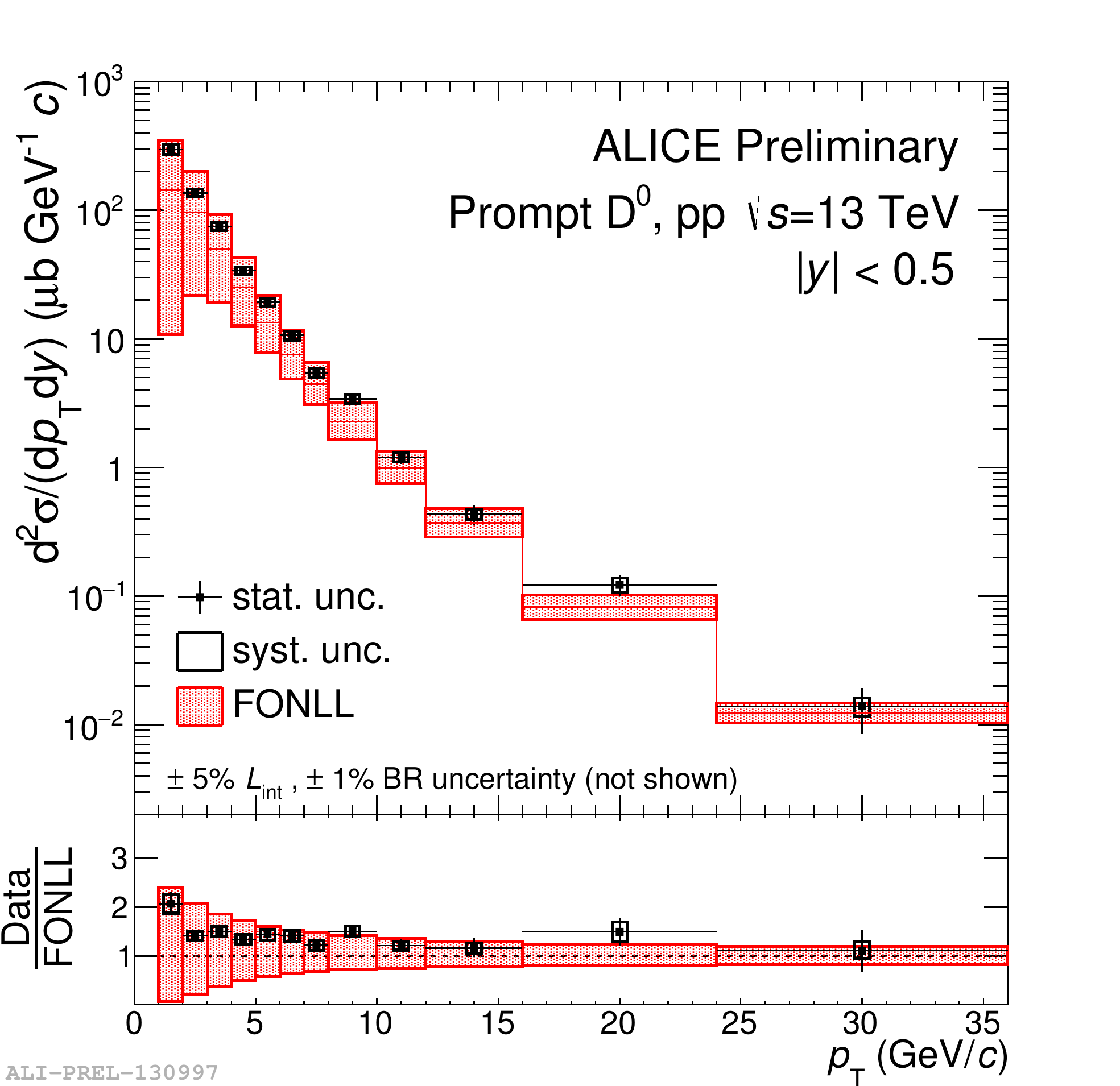}
        \caption{Preliminary $\pt$-differential production cross section of prompt $\Dzero$, measured by ALICE in pp collisions at $\sqrtS =$ 13 TeV,
        compared to FONLL prediction.}
        \label{fig:CrossSectionD0}
    \end{figure}

More recently, the measurement of the production cross sections of prompt $\Dzero$, $\Dplus$ and $\Dstar$
have been performed at \mbox{$\sqrtS=5$ TeV} and at the highest available LHC energy, \mbox{$\sqrtS=13$ TeV}.
The preliminary results for $\Dstar$ in pp collisions at 5 TeV and for $\Dzero$ at 13 TeV are presented
Fig.~\ref{fig:CrossSectionDstar} and~\ref{fig:CrossSectionD0} respectively.
Perturbative QCD calculations at NLO, carried out within the FONLL and GM-VFNS frameworks,
describe the measured D-meson cross-sections, within uncertainties,
over a large energy range.
Interestingly, such theoretical models manage to describe at the same time
the whole $\pt$ range of the measurement,
though theoretical uncertainties are quite large at low $\pt$.
In particular, FONLL is compatible with the $\pt$-differential production cross section of prompt $\Dzero$
from $\pt=0$ to 36 $\gmom$
\cite{alice2017dmesonProductionInppAt7TeVpass4,alice2016DmesonProductionInpPb5pt02TeVAndppAt7TeV}.

Furthermore, ALICE has evaluated the total charm cross-section in minimum-bias pp collisions at $\sqrtS=7$ TeV,
by extrapolating the measured $\Dzero$ production cross section to the full phase space
and compensating fragmentation contribution of charm quarks into $\Dzero$ mesons
\cite{alice2017dmesonProductionInppAt7TeVpass4,alice2016DmesonProductionInpPb5pt02TeVAndppAt7TeV}.
The resulting excitation function of the $c\bar{c}$ cross section,
taking into account measurements from other experiments,
is in agreement with pQCD calculations at NLO (MNR \cite{mnr1992totalCharmCrossSectionAtNLO}),
though data are at the upper edge of the theory uncertainty band.
It is worth mentioning that the MNR calculation is consistent with data over more than three orders of magnitude in $\sqrtS$.
All in all, pQCD calculations of production cross sections of hadrons containing a heavy quark
usually have large systematic uncertainties, compared to measurements, mainly driven by scale uncertainties.

\begin{figure}[htb]
        \includegraphics[scale=0.38]{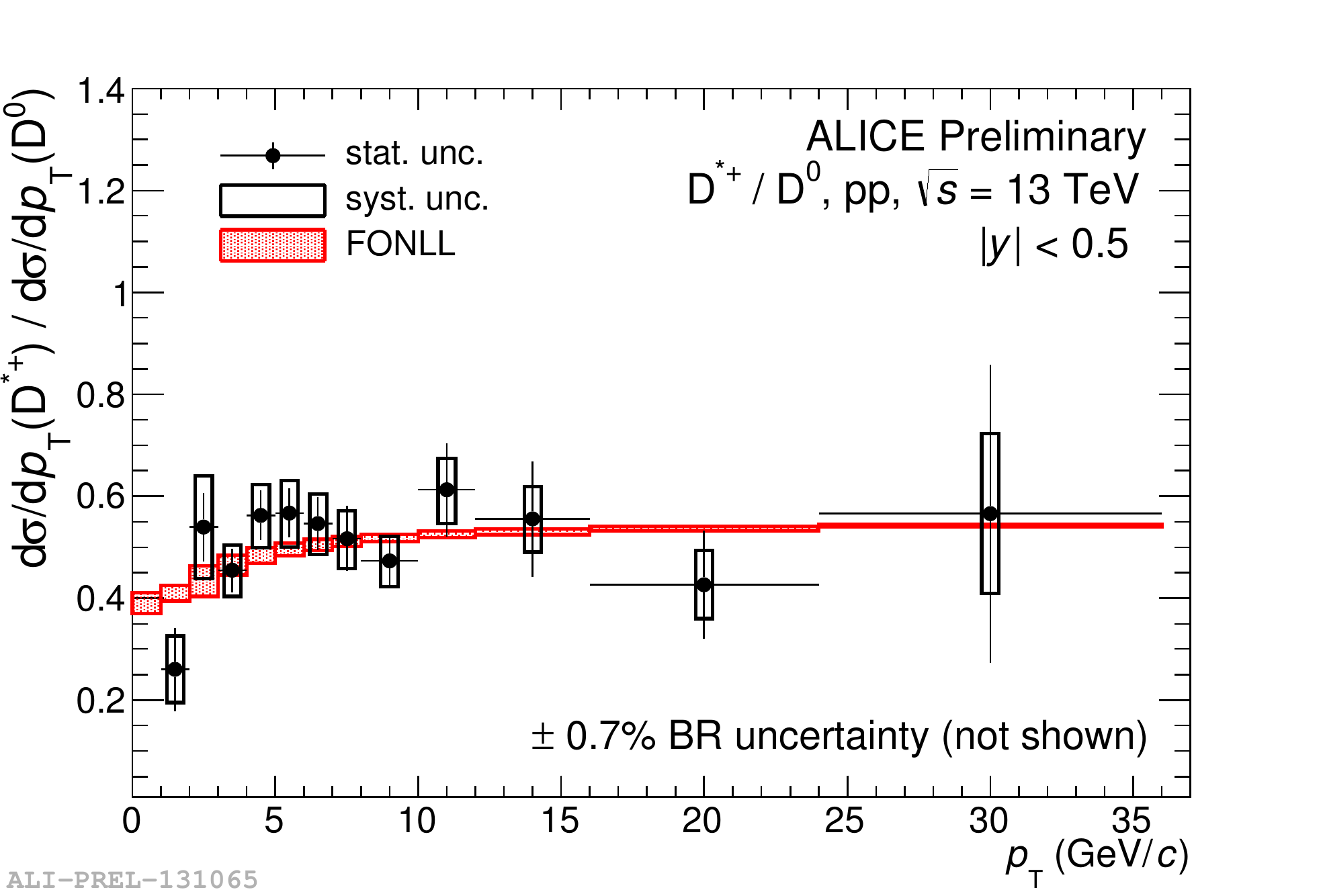}
        \caption{Preliminary comparison of the $\Dstar$ and $\Dzero$ production cross section in minimum-bias pp collisions at $\sqrtS=13$ TeV,
        superimpose to the FONLL prediction.}
        \label{fig:Ratio13DstarDzero}
    \end{figure}

\begin{figure}[htb]
        \includegraphics[scale=0.38]{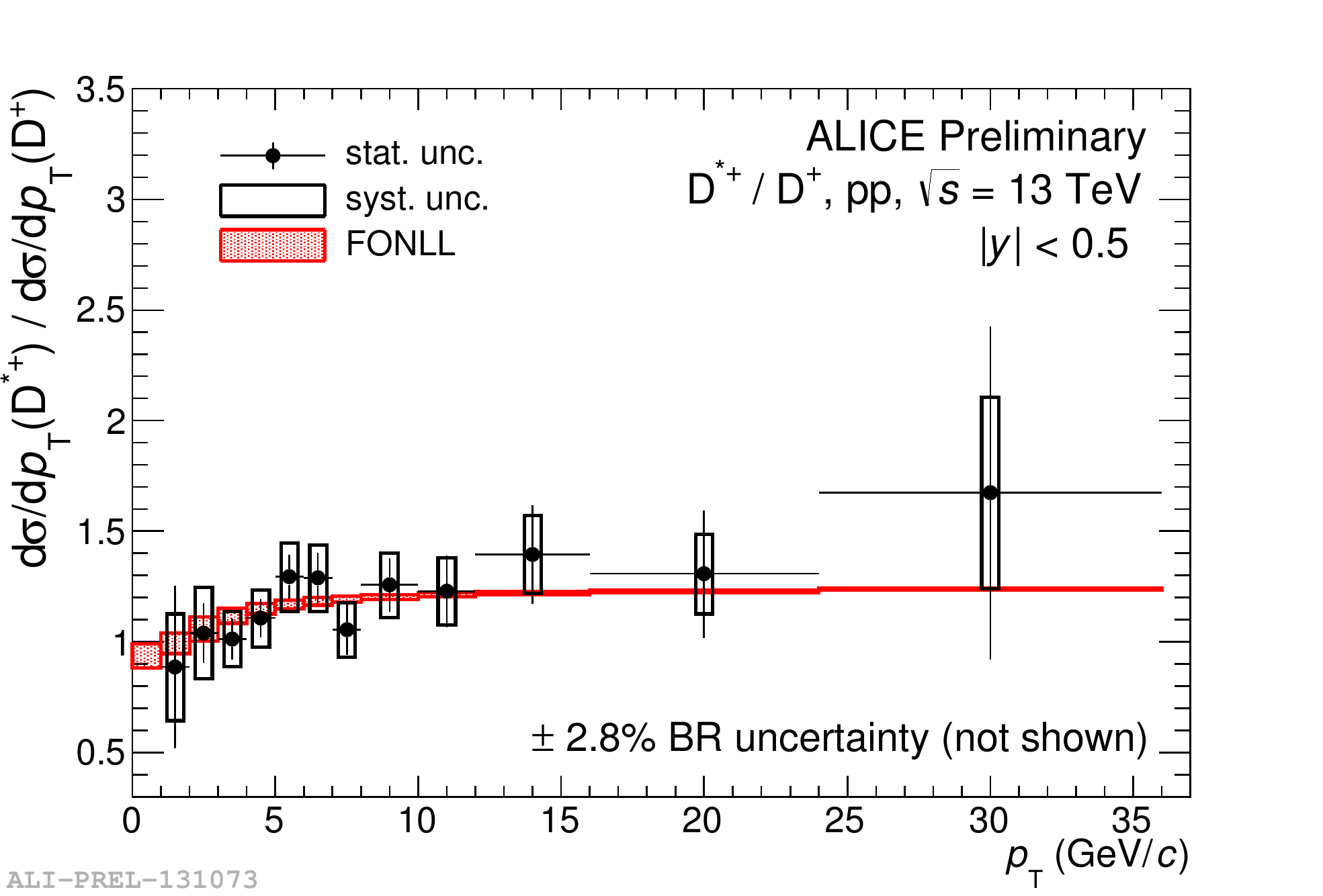}
        \caption{Preliminary comparison of the $\Dstar$ and $\Dplus$ production cross section in minimum-bias pp collisions at $\sqrtS=13$ TeV,
        superimpose to the FONLL prediction.}
        \label{fig:Ratio13DstarDplus}
    \end{figure}

\begin{figure}[htb]
        \includegraphics[scale=0.38]{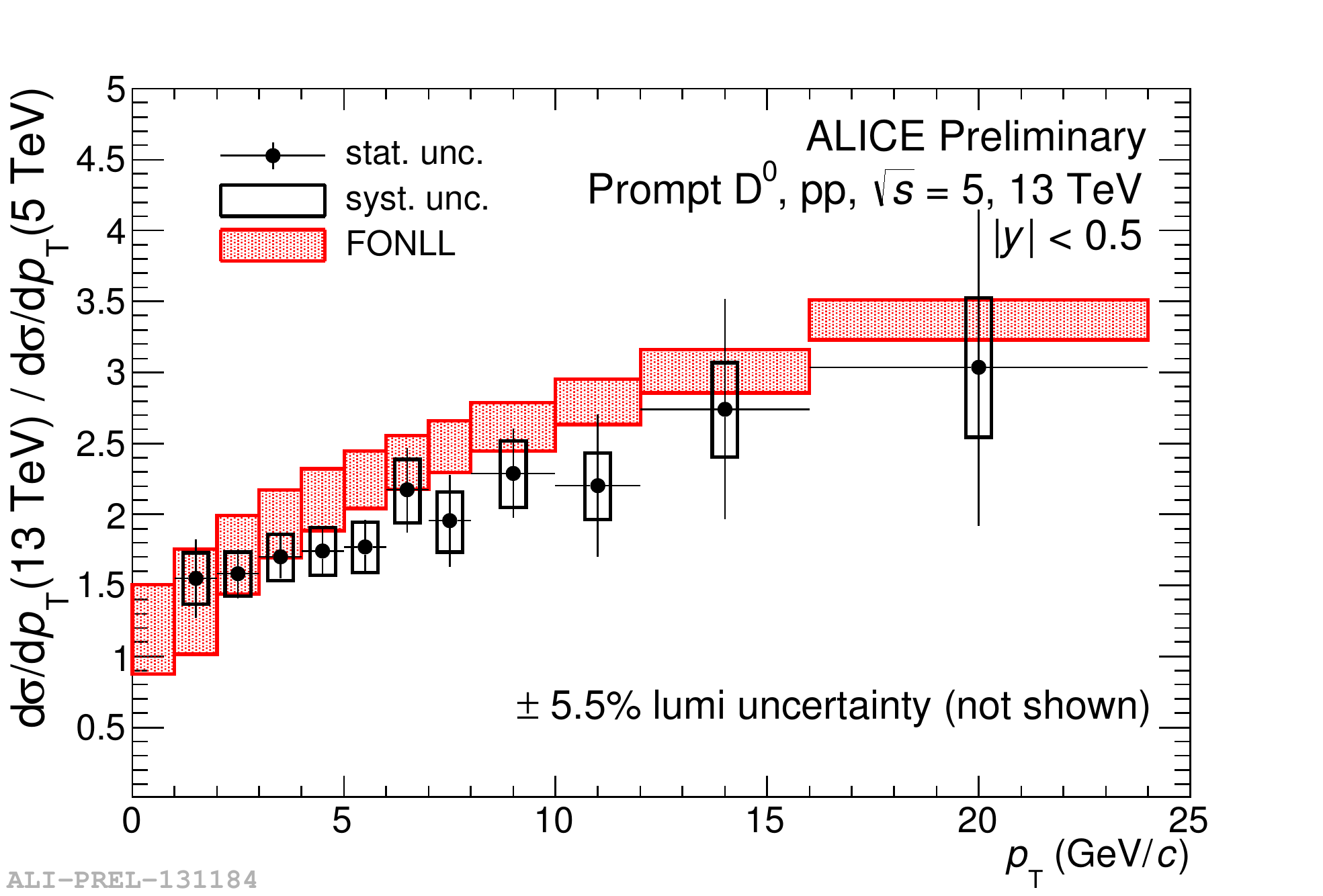}
        \caption{Preliminary ratio of the $\Dzero$ production cross sections in minimum-bias proton-proton collisions at $\sqrtS=$ 5 and 13 TeV,
        compared to FONLL prediction.}
        \label{fig:Ratio13over8D0}
    \end{figure}

The species dependence of the production cross sections of heavy-flavour hadrons
has been studied in minimum-bias pp collisions at \mbox{$\sqrtS=7$ TeV} by ALICE
\cite{alice2017dmesonProductionInppAt7TeVpass4}.
Such a dependence is studied in terms of cross-section ratios amongst D-meson species.
Preliminary results are also available at LHC Run 2 energy,
as shown Fig.~\ref{fig:Ratio13DstarDzero} and \ref{fig:Ratio13DstarDplus},
where the production cross sections, in minimum-bias \mbox{$\sqrtS=13$ TeV} pp collisions,
of $\Dstar$ and $\Dzero$ and of $\Dstar$ and $\Dplus$ are respectively compared.
Similarly, ratios of D-meson production cross sections at two different LHC energies
have been tackled to investigate the energy dependence of the heavy-flavour hadron productions.
The preliminary measurements of the ratio of the $\Dzero$ production cross sections at mid-rapidity
in pp collisions at 5 and 13 TeV are presented Fig.~\ref{fig:Ratio13over8D0}.
An overall agreement of measurements with FONLL calculations is observed.
It can be noted that the FONLL systematic uncertainties are significantly reduced
due to the independence of the renormalisation and factorisation scales
with respect to beam energies and meson species \cite{cacciari2015gluonPDFconstraintsFromRatioVsRapidityVsEnergy}.

    \section{Multiplicity dependence of heavy-flavour hadron yields in pp collisions}
    \label{sec:MultDepCrossSection}

\begin{figure}[htb]
        \includegraphics[scale=0.38]{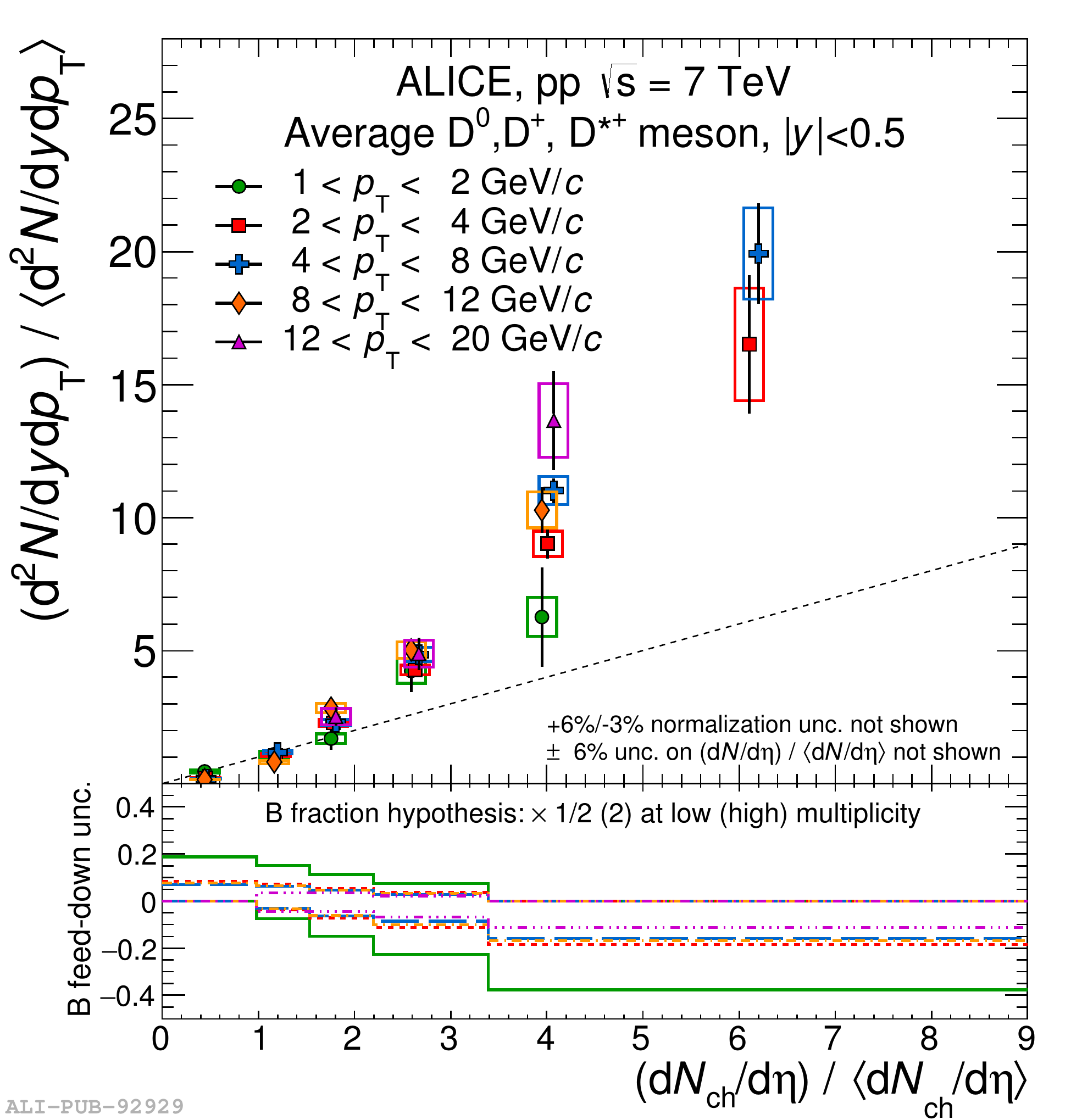}
        \caption{Average of $\Dzero$, $\Dplus$, $\Dstar$ relative yield as a function of relative charged-particle multiplicity, at central rapidity
        \cite{alice2015charmBeautyVSmultiplicityPP7TeV}.}
        \label{fig:DvsMult7TeV}
    \end{figure}

The event charged-particle multiplicity dependence of the $\Dzero$, $\Dplus$ and $\Dstar$ yields,
normalised to their respective multiplicity-integrated yields,
has been studied by the ALICE experiment in pp collisions at $\sqrtS=7$ TeV
\cite{alice2015charmBeautyVSmultiplicityPP7TeV}.
A steeper-than-linear increase of yields with the particle multiplicities in events,
similar for all the measured D mesons, has been observed.
Indeed the yields are enhanced by almost a factor of 15, relative to the minimum-bias yield,
for events with multiplicity six times higher than average inelastic-collision multiplicities,
as can be observed in Fig.~\ref{fig:DvsMult7TeV}.
Similar enhancements are observed for prompt $\Jpsi$, as well as for $\Jpsi$ coming from the decay
of beauty hadrons.
The comparison of open- and hidden-charm production implies that
the observed yield increase is likely related to $c\bar{c}$ and $b\bar{b}$ production processes,
and not strongly influenced by hadronisation.
However it should be noticed that a direct comparison has to be taken with care,
since the $\eta$ and $\pt$ regions of D mesons and $\Jpsi$ measurements are different.

The evolution of the charmed-meson yields with the event activity
is foretold by calculations including multi-parton interactions (MPI) at LHC energies.
The PYTHIA8 model \cite{torbjorn2008pythia8},
where several 2$\rightarrow$2 hard processes can take place in a given pp collision,
predicts an almost linear increase of D-meson yields with event particle multiplicities.
The percolation model,
assuming exchanges of colour sources with finite spatial extension as driving the hadronic collisions,
predicts a steeper-than-linear yield enhancement of D-mesons with event multiplicity \cite{ferreiro2015openCharmPercolation}.
Finally EPOS3, based on the Gribov-Regge formalism,
also anticipates a steeper-than-linear increase driven by a hydrodynamic evolution of the pp collision \cite{werner2014epos3}.
However, the D-meson yields are underestimated by these three models
when the event multiplicity exceeds two times the average inelastic-collision multiplicities,
even though calculations reasonably reproduce the data below such event activity.
Even if MPI seem to be an important input for hadronic collision event generators at high multiplicity,
one should keep in mind that alternative explanations may account for the charmed-meson yields enhancement.
For instance, the growth of gluon radiations associated to hard processes could explain the observed yield increase.
Nevertheless, no implementation of such effects is currently available in models
\cite{alice2015charmBeautyVSmultiplicityPP7TeV}.

    \section{Conclusion}
    \label{sec:Conclusion}

Heavy quarks are key tools to probe the pQCD and its interplay
with softer phenomena from the surrounding collectivity.
By exploiting LHC Run~1 proton-proton collisions, ALICE has constrained pQCD-inspired models
by means of various precise measurements of D mesons.
Calculations describe fairly well the production cross sections of open-charm meson
over wide $\pt$ ranges and at LHC energies ranging from $\sqrtS=2.76$ to 13 TeV.
More differential measurements, as a function of the event multiplicity for instance,
allow testing phenomenological models depicting a global view of hadronic collisions.
However, the description of softer QCD phenomena remain challenging for
hadronic collision event generators.

The increase of the LHC collision energy, for the second data taking period,
allows for more precise measurements.
Thus, QCD processes occurring at TeV collision energies will further be investigated.
Finally, the recent first measurements of open-charm baryons, $\mathrm{\Lambda_{c}^{+}}$ and $\mathrm{\Xi_{c}^{+}}$,
by ALICE in pp collisions is an additional input for \emph{ab initio} and phenomenological models.

\nocite{*}
\bibliographystyle{elsarticle-num}
\bibliography{thebibliography}

\begin{thebibliography}{10}
\expandafter\ifx\csname url\endcsname\relax
  \def\url#1{\texttt{#1}}\fi
\expandafter\ifx\csname urlprefix\endcsname\relax\def\urlprefix{URL }\fi
\expandafter\ifx\csname href\endcsname\relax
  \def\href#1#2{#2} \def\path#1{#1}\fi

\bibitem{fonll2012CharmAndBottomProductionAtLHC}
M.~Cacciari, S.~Frixione, N.~Houdeau, M.~M. L., P.~Nason, G.~Ridolfi,
  {Theoretical predictions for charm and bottom production at the LHC}, JHEP 10
  (2012) 137.
\newblock \href {http://arxiv.org/abs/1205.6344} {\path{arXiv:1205.6344}}.

\bibitem{gmvfns2012CharmProductionAtLHC}
B.~A. Kniehl, G.~Kramer, I.~Schienbein, H.~Spiesberger, {Inclusive
  Charmed-Meson Production at the CERN LHC}, Eur. Phys. J. 72 (2012) 2082.
\newblock \href {http://arxiv.org/abs/1202.0439} {\path{arXiv:1202.0439}}.

\bibitem{kTfactorisation2013CharmProductionAtLHC}
R.~Maciula, A.~Szczurek, {Open charm production at the LHC -
  $k_{t}$-factorization approach}, Phys. Rev. 87~(9) (2013) 094022.
\newblock \href {http://arxiv.org/abs/1301.3033} {\path{arXiv:1301.3033}}.

\bibitem{cacciari2015gluonPDFconstraintsFromRatioVsRapidityVsEnergy}
M.~Cacciari, M.~L. Mangano, P.~Nason, {Gluon PDF constraints from the ratio of
  forward heavy-quark production at the LHC at $\sqrt{S}=7$ and 13 TeV}, Eur.
  Phys. J. C75~(12) (2015) 610.
\newblock \href {http://arxiv.org/abs/1507.06197} {\path{arXiv:1507.06197}}.

\bibitem{alice2008pstationExpmtInJINST}
{ALICE Collaboration}, {The ALICE experiment at the CERN LHC}, JINST 3 (2008)
  S08002.
\newblock \href {http://dx.doi.org/10.1088/1748-0221/3/08/S08002}
  {\path{doi:10.1088/1748-0221/3/08/S08002}}.

\bibitem{alice2014AlicePerformanceRunI}
{ALICE Collaboration}, {Performance of the ALICE Experiment at the CERN LHC},
  Int. J. Mod. Phys. A29 (2014) 1430044.
\newblock \href {http://arxiv.org/abs/1402.4476} {\path{arXiv:1402.4476}}.

\bibitem{particleDataGroup2016}
C.~Patrignani, et~al., {Review of Particle Physics}, Chin. Phys. C40~(10)
  (2016) 100001.
\newblock \href {http://dx.doi.org/10.1088/1674-1137/40/10/100001}
  {\path{doi:10.1088/1674-1137/40/10/100001}}.

\bibitem{alice2012charmProductionInPP7TeV}
{ALICE Collaboration}, {Measurement of charm production at central rapidity in
  proton-proton collisions at $\sqrt{s} = 7$ TeV}, JHEP 01 (2012) 128.
\newblock \href {http://arxiv.org/abs/1111.1553} {\path{arXiv:1111.1553}}.

\bibitem{alice2017dmesonProductionInppAt7TeVpass4}
{ALICE Collaboration}, {Measurement of D-meson production at mid-rapidity in pp
  collisions at $\mathbf{\sqrt{s}=7}$ TeV. }\href
  {http://arxiv.org/abs/1702.00766} {\path{arXiv:1702.00766}}.

\bibitem{alice2012charmProdInPP2pt76TeV}
{ALICE Collaboration}, {Measurement of charm production at central rapidity in
  proton-proton collisions at $\sqrt{s}=2.76$ TeV}, JHEP 07 (2012) 191.
\newblock \href {http://arxiv.org/abs/1205.4007} {\path{arXiv:1205.4007}}.

\bibitem{alice2016DmesonProductionInpPb5pt02TeVAndppAt7TeV}
{ALICE Collaboration}, {D-meson production in p-Pb collisions at $\sqrt{s_{\rm
  NN}}=5.02$ TeV and in pp collisions at $\sqrt{s}=7$ TeV}, Phys. Rev. C 94~(5)
  (2016) 054908.
\newblock \href {http://arxiv.org/abs/1605.07569} {\path{arXiv:1605.07569}}.

\bibitem{sharma2017pp8TeV}
{A. Sharma for the ALICE collaboration}, {Measurement of D-meson production in
  pp collisions with ALICE at the LHC}, in: {22nd DAE-BRNS High Energy Physics
  Symposium Delhi, India, December 12-16, 2016}, 2017.
\newblock \href {http://arxiv.org/abs/1705.05147} {\path{arXiv:1705.05147}}.

\bibitem{mnr1992totalCharmCrossSectionAtNLO}
M.~L. Mangano, P.~Nason, G.~Ridolfi, {Heavy quark correlations in hadron
  collisions at next-to-leading order}, Nucl. Phys. B373 (1992) 295--345.
\newblock \href {http://dx.doi.org/10.1016/0550-3213(92)90435-E}
  {\path{doi:10.1016/0550-3213(92)90435-E}}.

\bibitem{alice2015charmBeautyVSmultiplicityPP7TeV}
{ALICE Collaboration}, {Measurement of charm and beauty production at central
  rapidity versus charged-particle multiplicity in proton-proton collisions at
  $ \sqrt{s}=7 $ TeV}, JHEP 09 (2015) 148.
\newblock \href {http://arxiv.org/abs/1505.00664} {\path{arXiv:1505.00664}}.

\bibitem{torbjorn2008pythia8}
T.~Sjostrand, S.~Mrenna, P.~Z. Skands, {A Brief Introduction to PYTHIA 8.1},
  Comput. Phys. Commun. 178 (2008) 852--867.
\newblock \href {http://arxiv.org/abs/0710.3820} {\path{arXiv:0710.3820}}.

\bibitem{ferreiro2015openCharmPercolation}
E.~G. Ferreiro, C.~Pajares, {Open charm production in high multiplicity
  proton-proton events at the LHC. }\href {http://arxiv.org/abs/1501.03381}
  {\path{arXiv:1501.03381}}.

\bibitem{werner2014epos3}
K.~Werner, B.~Guiot, I.~Karpenko, T.~Pierog, {Analysing radial flow features in
  p-Pb and p-p collisions at several TeV by studying identified particle
  production in EPOS3}, Phys. Rev. C89~(6) (2014) 064903.
\newblock \href {http://arxiv.org/abs/1312.1233} {\path{arXiv:1312.1233}}.

\end{thebibliography}







\end{document}